\documentclass[10pt]{article}
\usepackage{amsmath,amsfonts,amsthm,amssymb,amscd}
\usepackage[toc,page]{appendix}
\numberwithin{equation}{section}

\newcommand{\be}{\begin{equation}}
\newcommand{\ee}{\end{equation}}
\newcommand{\bs}{\begin{split}}
\newcommand{\es}{\end{split}}
\newcommand{\ba}{\begin{align}}
\newcommand{\ea}{\end{align}}
\newcommand{\basl}[1]{\begin{align}\begin{split}\label{#1}}
\newcommand{\bas}{\begin{align}\begin{split}}

\newtheorem{theo}{Theorem}[section]
\newtheorem{prop}[theo]{Proposition}
\newtheorem{lemm}[theo]{Lemma}

\newcommand\fpr{\hfill$\Box$\null}

\newcommand\R{\mathbb{R}}
\newcommand\C{\mathbb{C}}

\title{Ground state photon number at large distance}

\author{ L. Amour, L. Jager, J. Nourrigat}

\date{}

\begin{document}

\maketitle

\begin{abstract}
\noindent The purpose of this article is to give a result of localization in space of the ground state photons, in some sense,  of a Hamiltonian modelling nuclear magnetic resonance in  quantum electrodynamics. The asymptotic at infinity obtained is $|x|^{-{5}}$ where $x$ is considered here as the position of the photons. Moreover, the number of photons at large distance is the smallest in the ground state total spin direction.
\end{abstract}


\parindent=0pt
\parindent = 0 cm

\parskip 10pt
\baselineskip 12pt

\section{Introduction.}\label{s1}

This article is concerned with the ground states of the Hamiltonian modelling nuclear
magnetic resonance (NMR) in the framework of quantum electrodynamics (QED), in particular with
the distribution in $\R^3$  of photons in these states. 

NMR is the interaction of a finite number of fixed particles of $\R^3$  with a
constant magnetic field and with  the quantized field. This phenomenon is represented by a
Hamiltonian $H(g)$ depending on a positive parameter $g$ (the coupling constant) and
acting on a Hilbert space $\cal H$. This Hamiltonian, introduced by Reuse
\cite{Reu},  may also be seen as a simplification of the Pauli-Fierz Hamiltonian
\cite{BFS,F-73}. It is recalled in
Section 2.

The Hilbert space $\cal H$ is a completed tensor product
${\cal H}=  {\cal H}_{ph} \otimes  {\cal H}_{sp}$, where  ${\cal H}_{ph} $ is the space of
the photons and ${\cal H}_{sp} $, the space of the  particles with spin (cf. Section 2).
The space  ${\cal H}_{ph} $ is a Fock space, in which the usual operators, as the
number operator $N$ or the annihilation operators $a(k),k\in \R^3$, are defined.
 
It was shown in  \cite{BFS,G-polaron,S-1989,C-G,Gross-1}$,\dots$
that the infimum of
the spectrum is an eigenvalue. This result is recalled in Theorem \ref{t-gr-state}.
The eigenvectors are called ground states. Their properties are studied in, e.g.,
 \cite{H-S-1,A-H,HS-RD,DRGK15,SP,A-N-CS}.
The main result, Theorem \ref{asym-faible}, is concerned with the localization in space
of the photons of the ground state, for a non zero external field and a positive, but
sufficiently small, coupling constant $g$.

The problem of the localization of photons is mentioned in Reuse \cite{Reu}
(page 296-297) and  \cite{F-S}.
See   \cite{J-P} and  \cite{A} as well.

We first define a function on $\R^3$,  which  could be thought of as  the density
in space of the numbers of photons.
It is well known (see  \cite{DRGK15,H03}, see also \cite{G-polaron} and \cite{BFS,C-G}) that  every ground state $U$ of the Hamiltonian
$H(g)$ is in the domain $D( N^{m} \otimes I)$ for an arbitrary $m\geq 0$, where $N$ is the
number operator. For any $U$ in $D( N^{1/2} \otimes I)$, one recalls in the Appendix A the
definition of the function $k \rightarrow (a(k)\otimes I) U  $   an an element of
$L^2(\R^3 , {\cal H} ^3)$.

Therefore one may define, for any ground state $U$, the function  $k \rightarrow
(a(k)\otimes I) U $  in  $L^2(\R^3 , {\cal H} ^3)$, and hence its Fourier transform, which is
in $L^2(\R^3 , {\cal H} ^3)$ too. We denote this transform by
$x \rightarrow ( \widehat a(x )\otimes I)  U$. 

By abuse of notation, one writes:
%---
\be\label{ann-x} (\widehat a(x )\otimes I) U = \int _{\R^3}  e^{-i x\cdot k }
(a(k)\otimes I) U  d k. \ee
%------
One has:
%---
$$ (2\pi )^{-3}  \int _{\R^3} \Vert (\widehat a(x ) \otimes I) U  \Vert ^2 dx =
< (N \otimes I) U, U>. $$
%---
Since  $< (N \otimes I) U, U>$  is the average number of photons, one could imagine
that the average number of photons in a Borel set $E$ of $\R^3$ is given by:
%---
$$ < N_Ef , f> = (2\pi )^{-3}  \int _{E} \Vert \widehat a(x ) \otimes I) U  \Vert ^2 dx  . $$
%----
In this way, the function 
$ x  \rightarrow   \Vert (\widehat a(x ) \otimes I) U  \Vert ^2 $ can be seen as the density,
in space, of the number of photons.

We aim at studying this function when $U= U_g$ is a ground state of $H(g)$, under the
assumption that the constant field ${\bf B}^{ext}$ is not zero and that $g$ is positive
but sufficiently small. In this case, according to  \cite{S-1989} and  \cite{H}, the space of
the ground states has dimension $1$. This result is recalled in Theorem \ref{t-gr-state}.

The operator  $\sigma_m^{[\lambda]}$  is defined in (\ref{spin-ini}). The main result is the
following.
 
\begin{theo}\label{asym-faible} Suppose that   ${\bf B}^{ext} \not= 0$.  Let $U_g$  be a
normalized ground state of  Theorem  \ref{t-gr-state}, where the coupling constant $g$ is
small enough to ensure that the space of ground states has dimension $1$. Then:
\begin{enumerate}
\item
The function $x \rightarrow |x|^{5/2} (\widehat a(x ) \otimes I) U_g$ is bounded and
continuous on  $\R^3$, with values in ${\cal H}^3$.
\item
For every unitary vector $v$ of $\R^3$, one has:
%----
$$ \lim _{|x| \rightarrow \infty} |x|^{5/2} (\widehat a(|x|v  ) \otimes I)  U_g  =
 - \frac {3} {\sqrt {2}}   \chi (0)  ( v \times {\bf S} ^ {[tot]})  U_g  $$
%---
where  ${\bf S} ^ {[tot]}$  is the vector of $\R^3$ with coordinates:
%----
$$ S_j ^ {[tot]} = \sum _{\lambda = 1 } ^P< ( I \otimes \sigma_j^{[\lambda ]}  )  U _g, U _g >.$$
%---
\end{enumerate}
\end{theo}

In particular, for the local density of the photons of the ground state, there exists $C>0$
such that
%----
$$ \Vert (\widehat a(x )\otimes I)  U_g \Vert^2 \leq C ( 1 + |x|) ^{-5}. $$
%---
Since $U_g$ is normalized, one has, for all unitary vectors $v\in \R^3$:
%----
$$  \lim _{|x| \rightarrow \infty} |x|^{5} \Vert (\widehat a(|x|v  ) \otimes I)  U_g   \Vert ^2
=(9/2)  |\chi (0)|^2 |( v \times {\bf S}^ {[tot]} ) |^2 $$
%---
One sees that, for large distances,  the photons are fewer in the direction of the total spin.

 \section{The Hamiltonian and its ground state.}\label{s2}

 The Hilbert space of the states of our system is a completed tensor product
 ${\cal H}_{ph} \otimes {\cal H} _{sp}$, where  ${\cal H} _{ph}$ is the Hilbert space of the
 free photons and ${\cal H} _{sp}$, the space of the particles with spin.

{\it Photons.} The one photon configuration Hilbert  space ${\cal H}$  is the set of mappings
$f \in L^2 (\R^3, \R^3)$ satisfying $k\cdot f(k) = 0$ almost everywhere in $k\in\R^3$
(see \cite{L-L}) where $|f|^2=\int_{\R^3}|f(k)|^2dk$.  One denotes by $<f, g>$ the scalar
product of two elements $f$ and  $g$ of ${\cal H}$, where the mapping 
$g \rightarrow < f, g>$  is antilinear.  The Hilbert space ${\cal H}_{ph}$ of photon
quantum states is  the symmetrized Fock space
${\cal F}_s ({\cal H}_{\bf C})$ over  the complexified space of ${\cal H}$.
We follow \cite{RSII} for Fock spaces considerations and notations, in particular, for the
usual operators in these spaces: the Segal field $\Phi_S(V)$ associated with an element $V$ in
${\cal H}^2$, the $\Gamma (T)$ and ${\rm d} \Gamma (T)$ operators associated with some
operator $T$ acting in ${\cal H}^2$. Note that, throughout this paper, the space
${\cal H}^2$ is
sometimes identified to the complexified space ${\cal H}_{\bf C}$ but this identification is not
everywhere systematically effectuated in order to avoid possible confusions.

Let  $M_{\omega}$ be the operator with domain   $D(M_{\omega}) \subset {\cal H}$ such
that $M_{\omega} q (k) = |k| q(k)$ almost everywhere in $k\in\R^3$.
We denote in the same way the analogous operators defined  on ${\cal H}^2$ or
on the complexified  space ${\cal H}_{\bf C}$. In the Fock space framework, the photon free
energy Hamiltonian operator
$H_{ph}$  is usually defined as  $  H_{ph} =  {\rm d} \Gamma (M_{\omega})$.

The photon number operator denoted by $N$ is $N = {\rm d} \Gamma (I)$.

The three components of the magnetic field at each point $x$ in  $\R^3$ are defined
using the elements $ B_{jx}$ belonging to  ${\cal H}^2$ and written as
follows,  when one identifies ${\cal H}^2$ with the complexified space ${\cal H}_{\bf C}$:
 %--------
\be\label{7.3} B_{jx}(k) = {i\chi(|k|)|k|^{1\over 2} \over (2\pi)^{3\over 2}}
e^{-i(   k\cdot x   )} {k\times e_j \over |k|},\quad k\in\R^3\backslash\{0\}\ee
 %-----
where the function $\chi $ (ultraviolet cutoff) belongs to ${\cal S} (\R)$.

One then defines the  magnetic fields components operators at each point
 $x$ of $\R^3$ by:
%----
$$B_m(x) = \Phi_{S } (B_{mx}),$$
%---
for $m=1,2,3$.

{\it Spins.}  The configuration space of the system of $P$ spins is then the space
${\cal H}_{sp} =( \C^2  )^{\otimes P} $.
The fermion property for the spin-$\frac{1}{2}$ fixed particles is omitted here.
In the space  ${\cal H}_{sp} $, we use the operators related to the
spins of the different particles. Let  $\sigma _j$ ($1 \leq j \leq 3$)
be the Pauli matrices:
%----
\be\label{Pauli} \sigma_1 = \begin{pmatrix}  0 & 1 \\ 1 & 0   \end{pmatrix},
\qquad
\sigma_2 = \begin{pmatrix} 0 & -i \\ i & 0   \end{pmatrix},
\qquad
\sigma_3 = \begin{pmatrix}  1 & 0 \\ 0 & -1  \end{pmatrix}.\ee
%----
For all $\lambda \leq P$ and all  $m\leq 3$, we denote by $\sigma_m^{[\lambda]}$
the operator in ${\cal H} _{sp}$ defined by:
%---
\be\label{spin-ini}\sigma_m^{[\lambda]} = I \otimes \cdots  \otimes I \otimes \sigma_m\otimes  I  \otimes \cdots \otimes I,
\ee
%---
where  $\sigma_m$ is located at the  $\lambda ^{th}$ position.

{\it The Hamiltonian.}  This Hamiltonian is often used for modeling NMR in quantum field
theory (see  \cite{Ro-Au,J-H} and  Section 4.11 of \cite{Reu}).  It is a selfadjoint
extension of the following operator, initially defined  in a dense subspace of
${\cal H}_{ph} \otimes {\cal H}_{sp} $:
%---
\be\label{7.1} H(g) = H _0 + g   H_{int}   ,\ee
%---
where $g$ is a positive constant and:
%---
\be\label{7.2}     H_0 = H_{ph} \otimes I +  \sum _{\lambda =1}^P  \sum _{m=1}^3
B^{ext}  _m \otimes
\sigma_m^{[\lambda]}\ee
%----
where $H_{ph} = {\rm d} \Gamma (M_{\omega})$ is the photon free energy operator, acting in
a domain $D(H_{ph}) \subset {\cal H}_{ph}$ and
${\bf B}^{ext}  = (B^{ext} _1 ,  B^{ext} _2 , B^{ext}_3) \not = 0$
is the constant magnetic field. Moreover:
%----
 \be\label{interact}  H_{int}  =  \sum _{\lambda =1}^P \sum _{m=1}^3
B  _m (x_{\lambda}) \otimes  \sigma_m^{[\lambda]}\ee
%----
 and the  $x_{\lambda }$ ($1\leq \lambda \leq N$)  are the points of
$\R^3$ where the fixed particles are located.

If an element $U$ of ${\cal H}^2$  lies in the domain  $ D( M_{\omega }^{-1/2})$
then the Segal  field  $\Phi_S(U)$ is bounded from  $D(H_{ph})$ into ${\cal H}_{ph}$,
see point ii) of Proposition 3.4 in
\cite{A-L-N-2}  or see {\cite{DG}}.
This is therefore the case for the operators $B_j (x)$ and $E_j(x)$ according
to the assumptions on the ultraviolet cutoff function $\chi$ in  (\ref{7.3}).
Thus, according to the Kato-Rellich Theorem,   $H(h)$ has a selfadjoint
extension with the same domain as the free operator $H_0 = H_{ph } \otimes I$ domain.

{\it The ground state.} Let us now recall the results of \cite{G-polaron}, \cite{S-1989}
and \cite{C-G}  about the ground state of this Hamiltonian, or more precisely of Hamiltonians
very close to this one. 

\begin{theo}\label{t-gr-state}
The operator $H(g)$ defined by (\ref{7.1}), (\ref{7.2}) and (\ref{interact}), admits a selfadjoint
extension with the same domain as the free operator $H_0 $.
There exists a unitary element  $U_g$ of $ D(H_0) $, such that $H(g)U_g = E(g)  U_g$, where
$E(g) $ is the infimum of the spectrum of  $H(g)$. This element  $U_g$ is in the domain of
$N^m \otimes I$, for all  $m\geq 0$. If $g$ is small enough and if
${\bf B}^{ext} \not= 0$, this element  $U_g$  is unique  up to a
multiplicative factor. 
\end{theo}

This theorem follows from  \cite{C-G} (Theorem 1 page 447), and, for the uniqueness, from 
\cite{S-1989} and  \cite{H}.

 \section{Asymptotic behavior at infinity  of the photon number.}\label{s4}

Theorem \ref{asym-faible} is proved in this Section.
 We first note that, according to  equality (\ref{expr-a(k)}), the mapping $k \rightarrow (a(k) \otimes I )U$ taking values in  ${\cal H}^3$, 
belongs to $L^1(\R^3)$, and consequently, its Fourier transform denoted by
$x \rightarrow (\widehat a(x )\otimes I) U $, also taking values in  ${\cal H}^3$, is continuous on  $\R^3$.

 The following result is useful in order to prove the other points in Theorem \ref{asym-faible}.

  \begin{prop}\label{th-sp}  Let $(H, D(H))$ be a self-adjoint operator in a Hilbert space ${\cal H}$ and suppose that  the spectrum of the operator $H$ is the half line
 $[E , \infty)$ (with  $E\in \R$). Then,   %---
  $$ F(z, H, E )  = z ( H-E + z )^{-1}$$
  %---
  is a holomorphic mapping on the half plane $\{ z\in {\bf C},  \,{\rm Re} z >0 \}$,  
  continuous on the closure of this half plane without the origin, and with a norm smaller than or equal to one.
Let $(z_n)$ be a sequence of complex numbers  satisfying ${\rm Re} z_n \geq 0 $ and converging to zero.
  Set $\varphi$ the function defined on the spectrum of  $H$ by $\varphi (E) = 1$ and
 $\varphi (x) = 0$ if $x>E$. Then, for any  $f\in {\cal H}$:
  %----
  $$ \lim _{n \rightarrow \infty }F(z_n, H, E)  f = \varphi (H)f = P f$$
  %---
where $P$ is the orthogonal projection on ${\rm ker} ( H - E)$.

  \end{prop}

  {\it Proof. } For any $x$ belonging to the spectrum of $H$, set:
  %----
  $$ \varphi_n (x) = \frac { z_n  } {z_n + x  -E }. $$
  %-----
If ${\rm Re} z_n \geq 0$ then $| \varphi_n (x) | \leq 1$ for all $x$ in the spectrum of $H$.
In addition, $\varphi _n (x)$ tends to $\varphi (x) $ for each $x$ in the spectrum of $H$. According to a standard result (for example, see
\cite{BS} Lemma 3, Chapter 5, Section 4),  $\varphi_n (H)f$ tends to
  $\varphi (H)f$ for each $f$. Observe that $\varphi_n (H) f = F(z_n, H, E)f$.
 It is also well known (see \cite{BS}, Theorem  3, Chapter 6,  Section 1 and see also \cite{PLB}) that $\varphi (H) $ is the
  orthogonal  projection on ${\rm ker} ( H - E)$.  The proof of the Proposition then follows.

\fpr

The mapping  $(\widehat a(x)\otimes I) U_g$ is given by (\ref{ann-x})
and  $( a(k)\otimes I)  U_g$ by (\ref{expr-a(k)}) and (\ref{7.3})  where $U_g$ is the ground state.
That is:
 %----
 $$ (\widehat a (x )\otimes I)   U_g = - \frac {ig} { 4  \pi^{3\over 2} }   \sum _{\lambda =1}^P
   \sum _{m=1}^3  \int _{\R^3}  e^{-i x\cdot k }  \chi(|k|)|k|^{1\over 2}
 {k\times e_m \over |k|} ( H-E + |k|) ^{-1} f_m^{[\lambda]}    dk $$
 %-----
with $f_m ^{[\lambda]} = (I  \otimes \sigma_m ^{[\lambda]}  )U_g$.
In the sequel,   $\chi (|k|)$ is approximated by
 $\chi (0) e^{-|k|} $. Therefore, the following function is under consideration:
 %---
 %----
 \be\label{b(x)}  b(x ) U_g = - \frac {ig} { 4  \pi^{3\over 2} }   \sum _{\lambda =1}^P
   \sum _{m=1}^3  \int _{\R^3}  e^{-i x\cdot k }  \chi(0) e^{-|k|} \ |k|^{1\over 2}
 {k\times e_m \over |k|} ( H-E + |k|) ^{-1} f_m ^{[\lambda]}   dk.\ee
 %-----

 \begin{lemm}\label{majo-err} There exists $C>0$ such that:
 %---
 $$ \Vert  (\widehat a (x )\otimes I)  U - b (x) U \Vert \leq \frac {Cg} {|x|^3}. $$
 %----

 \end{lemm}

 {\it Proof of the Lemma.} The standard measure on the unit sphere $S^2$ is denoted by $\mu$. For any $v\in S^2$, $m\leq 3$  and $\lambda >0$, one has:
 %---
 \be\label{int-sphere}  \int _{S^2} e^{-i\lambda v \cdot \omega } (\omega \times e_m)
 d \mu (\omega) = 4 i \pi  (v \times e_m) \left [ \frac { \cos \lambda } {\lambda }
 - \frac { \sin \lambda } {\lambda ^2} \right ]  .\ee
 %----
 One then deduces the existence of $C>0$ satisfying:
 %---
 $$ \Vert  (\widehat a (x )\otimes I)  U - b (x ) U \Vert \leq  C   \sum _{m=1}^3
 |I_m(x)| $$
 %---
with, if $x = |x|v$, $|v|= 1$:
 %---
 $$ I_m(x) = \int _0^{\infty } \left [ \frac { \cos |x| \rho   } {|x| \rho  }
 - \frac { \sin |x| \rho  } {(|x| \rho ) ^2} \right ]   \Phi_m (\rho)    d\rho   $$
 %------
where:
 %----
 $$  \Phi_m (\rho) =  \rho ^{5/2}  \Big (\chi(\rho ) - \chi(0)e^{-\rho} \Big )
  ( H-E + \rho ) ^{-1} f_m.
 $$
 %----
One checks that:
 %----
 $$  \int _0 ^{\infty }  \frac { \sin ( |x| \rho ) } { |x|^2 \rho^2 } (  \Phi_m (\rho))
 d\rho =
   \frac { 1 } { |x|^3 }  \int _0 ^{\infty } \cos ( |x| \rho )  \frac { d } { d \rho }
   \left [   \frac { \Phi_m (\rho) } { \rho ^2 } \right ] d \rho  $$
 %---
and:
%----
$$  \int _0 ^{\infty } \frac { \cos ( |x| \rho ) } { |x| \rho }
 (  \Phi_m (\rho))
 d\rho =   \frac { 1 } { |x|^3 }  \int _0 ^{\infty } \cos ( |x| \rho )  \frac { d^2 } { d \rho^2 }
  \left [   \frac {  \Phi_m (\rho) } { \rho  } \right ] d \rho. $$
 %----
These integrations by parts holds true since as $\rho\to 0$:%----
$$  \left | \frac { d } { d \rho }
   \left [   \frac {  \Phi_m (\rho) } { \rho ^2 } \right ] \right | +
  \left | \frac { d^2 } { d \rho^2 }
   \left [   \frac {  \Phi_m  (\rho) } { \rho  } \right ] \right |
   \leq   \frac { C } { \rho  ^{1/2}  }.$$
%----
Moreover, the left hand side is rapidly decreasing as $\rho\to \infty$. The proof is thus completed.  

 \fpr

 {\it Proof of Theorem  \ref{asym-faible}.  } It is sufficient to consider $b(x ) U_g$.
For each $v\in S^2$, one sees using  (\ref{b(x)}) and (\ref{int-sphere}):
 %----
 $$b (|x| v ) U_g = \chi (0)  \pi^{-1/2}   \sum _{\lambda =1}^P
  \sum _{m=1}^3  v \times e_m \ I_{m,aux}^{[\lambda]}  (|x|) $$
 %-----
with:
 %----
 $$I_{m,aux}^{[\lambda]}  (|x|) = \int _0^{\infty } \left [ \frac { \cos |x| \rho   } {|x| \rho  }
 - \frac { \sin |x| \rho  } {(|x| \rho ) ^2} \right ]   \Psi_m (\rho)    d\rho
  $$
 %------
where:
 %----
 $$  \Psi_m (\rho) =  \rho ^{5/2}   e^{-\rho}
  ( H-E + \rho ) ^{-1} f_m ^{[\lambda]}.
 $$
 %----
One has:
%---
\begin{equation}
  I_{m,aux}^{[\lambda]}  (|x|) = \frac{2}{|x|^{5/2}}\int_0^{\infty} (t^2 \cos(t^2) - \sin(t^2))e^{-t^2/|x|}
  F(\frac{t^2}{|x|},H,E) f_m ^{[\lambda]}  dt
\end{equation}
%------
where $F(s,H,E)=s(H-E+s)^{-1}$.

We can write $ I_{m,aux} ^{[\lambda]} = I^{\lambda +}_{m,aux}+ I^{\lambda -}_{m,aux}$ setting:
%----
\begin{equation}
  I^{\lambda \varepsilon }_{m,aux}= \frac{1}{|x|^{5/2}}\int_0^{\infty}e^{\varepsilon i t^2} (t^2+\varepsilon  i)
 e^{-t^2/|x|}
  F(\frac{t^2}{|x|},H,E) f_m ^{[\lambda]} \ dt
\end{equation}
with $\varepsilon =\pm$.
If $|x|\geq 1$, a change of contour of integration shows that:
%----
\begin{equation}
  I^{\lambda \varepsilon}_{m,aux}= \frac{1}{|x|^{5/2}}(\varepsilon  i e^{\varepsilon  i \pi /4})
  \int_0^{\infty} (r^2+1)e^{-r^2}\  e^{ - \varepsilon i r^2 /|x|}
  F(\frac{\varepsilon  i r^2}{|x|},H,E) f_m  ^{[\lambda]} \  dr.
\end{equation}
%------
It is therefore deduced that:
%-------
\begin{equation}
  |x|^{5/2} I_{m,aux} ^{[\lambda]}  (|x|) \leq 2 |f_m^{[\lambda]}  |  \int_0^{\infty} (r^2+1)e^{-r^2}\ dr.
\end{equation}
%---------
Consequently, the function
$ x\rightarrow |x|^{5/2} |b(x)|$ is bounded on $\{ |x| \geq 1 \}$ proving
 point i) of the Theorem  when using Lemma \ref{majo-err}.

Using Lebesgue's dominated convergence Theorem, point ii) comes from:
%----
$$ \lim _{|x| \rightarrow \infty } |x|^{5/2} I^{\lambda \varepsilon} _{m,aux} (|x|) =
(\varepsilon  i e^{\varepsilon  i \pi /4})
 \lim _{ z\rightarrow 0, {\rm Re }z \geq 0}  F (z , H, E)f_m^{[\lambda]}    \int_0^{\infty} (r^2+1)e^{-r^2} dr.$$
%---
The existence of the above limit comes from Proposition \ref{th-sp} which also shows:%----
$$ \lim _{ z\rightarrow 0, {\rm Re }z \geq 0}  F (z , H, E) f_m ^{[\lambda]} =
P f_m = < f_m^{[\lambda]} , U_g> U_g $$
where $P$ is the  projection on the eigenspace of the infimum of the spectrum of $H(g)$. Therefore:
%---
\begin{align*} \lim _{|x| \rightarrow \infty } |x|^{5/2}   b(|x| v ) U_g  &= - \sqrt {2/\pi}   \chi (0)
 \sum _{\lambda =1}^P  \sum _{m=1}^3  v \times e_m
 <(I  \otimes \sigma_m ^ {[\lambda]} )U_g , U_g> U_g  \int_0^{\infty} (r^2+1)e^{-r^2} dr \\
& = - \sqrt {2/\pi}   \chi (0) ( v \times {\bf S} ^ {[tot}  )  U_g  \int_0^{\infty} (r^2+1)e^{-r^2} dr\\
& = - \frac {3} {\sqrt {2}}  \chi (0)   ( v \times {\bf S} ) ^ {[tot}   U
\end{align*}
%------
proving point ii) with the help of Lemma \ref{majo-err}.

\fpr

\begin{appendices}

\section{Standard facts on annihilation operators. } 

The following results are classical but adapted to the Hilbert space 
 $H$ in Section \ref{s1}.
We denote by  ${\cal F}_s^{reg} ({\cal H}_{\bf C})$ the subspace of ${\cal F}_s ({\cal H}_{\bf C})$
constituted with the finite linear combinations of the symmetrized products of  $g_1 \otimes\cdots \otimes g_m$ where the $g_j\in {\cal S} (\R^3, \R^3)$ satisfy $k\cdot g_j(k) = 0$ for all $k\in\R^3$.

For all $k\in \R^3$ and  $f\in {\cal F}_s^{reg} ({\cal H}_{\bf C})$,
 $a(k)f$ is classically defined as following. For any
  (non symmetrized) product $g = g_1 \otimes\cdots \otimes g_m$,
with the  $g_j$ satisfying the above conditions, one set:
  %--
  $$ a(k)g = \sqrt {m} g_1(k) g_2 \otimes \cdots \otimes g_m.$$
  %---
This definition is next adapted to the symmetrized space.
Since the function $g_1$ takes values in  $\R^3$,
one sees that
$k\rightarrow a(k) f $ takes values in $( {\cal F}_s ({\cal H}_{\bf C}))^3$ and belongs to  ${\cal S} (\R^3)$.
One also notes:
%---
$$ <  Nf , f > =  \int _{\R^3} \Vert a(k) f \Vert ^2 dk. $$
%----

The next two propositions are useful for the purpose of an extension by density in $D(N^{1/2} )$.

\begin{prop}
   ${\cal S} (\R^3,\R^3) \cap {\cal H} $ is dense in ${\cal H} $.
\end{prop}

{\it  Proof. }
Set $f\in {\cal H}$ and $\rho$ a $C^{\infty}$ smooth cut-off function defined on $\R^+$, vanishing on $[0,1]$ and equal  to one  on  $[2,+\infty)$.
One has:
    $$
    \lim_{n\rightarrow \infty} \int_{\R^3} |f(k) - f(k)\rho(n|k|^2)|^2 \ dk=0
    $$
using Lebesgue Theorem.  The above norm  $| \ |$ is the one in  $\R^3$.
Thus, the set  ${\cal H}_0$ of functions in  ${\cal H}$ vanishing on a ball centered at the origin is dense in ${\cal H}$.
Take $f\in {\cal H} _0$ vanishing on a ball centered at $0$ with radius $\eta>0$.
 There exist sequences $f_i^{(n)} $  of
     ${\cal S} (\R^3,\R)$ converging to $f_i$,  for $i=1,2,3$ 
    ($n$ is the coordinate index   for these sequences). It is naturally not clear that:
    $$
\sum_1^3 k_i f_i^{(n)}(k) = 0
$$
and therefore that $ f^{(n)} $ (whose components  are the $ f_i^{(n)}$) belongs to ${\cal H}$.
We then set, again with a cut-off function:
%----
\be
g^{(n)}= (f^{(n)} -\frac{k}{|k|^2} f^{(n)}\cdot k)\
\rho(\frac{4}{\eta^2}|k|^2) .
\ee
%------
One checks that $g^{(n)}\in {\cal H} \cap {\cal S}(\R^3,\R^3)$
and tends to $f$ in ${\cal H}$, i.e., with the 
$ L^2(\R^3,\R^3) $ norm.

\fpr

\begin{prop}
   ${\cal F}_s^{reg}({\cal H}_{\C})$ is dense in  $D(N^{1/2} )$.
\end{prop}

{\it Proof. } One chooses a basis $(u_i)$ of ${\cal H}$ in
${\cal H} \cap {\cal S}(\R^3,\R^3)$ and set:
%----
\begin{equation}\label{zeta-alpha}
  \zeta_{\alpha}= \sqrt{\frac{|\alpha| ! }{\alpha !}}
  S_n\left(\underset{i}{\otimes}\, u_i^{\otimes \alpha_i}\right),\quad n=|\alpha|.
\end{equation}
%------
The set of  $ \zeta_{\alpha}$ (with  $|\alpha|=n$) is an othonormal basis of
$\otimes^n_s{\cal H} $ (see Janson \cite{Jan}).
Thus, the set of finite linear combinations of $ \zeta_{\alpha}$ (with arbitrary 
$|\alpha|$) is dense in the set of finite number particles 
${\cal F}^{fin}({\cal H} )$ which is dense in 
$D(N^{1/2} )$. The proof of the Proposition is then complete since the set of finite linear combinations of  $ \zeta_{\alpha}$ is included in ${\cal F}_s^{reg}({\cal H}_{\C})$.

\fpr

We can now extend the definition  
of  $a(k)$ to $D(N^{1/2} )$.

\begin{prop}\label{a-k}
Set $f\in D(N^{1/2} )$ and a sequence $(f_n)$ in 
 ${\cal F}_s^{reg} ({\cal H}_{\bf C})$  converging to $f$ in $D(N^{1/2} )$.
Then, the  $k \rightarrow a(k) f_n$
has a limit in $L^2 (\R^3, ( {\cal F}_s ({\cal H}_{\bf C}))^3)$.
This limit is denoted  (definition) by $a(k)f$.
\end{prop}

Indeed, according to the previous points, one has if $m <n$:
 %----
 $$ \int _{\R^3} \Vert a(k) (f_m - f_n)  \Vert ^2 dk \leq
 ( \Vert N^{1/2}  f_m \Vert +  \Vert N^{1/2}  f_n \Vert )
 \Vert N^{1/2} ( f_m - f_n )\Vert$$
and one also has ($\varphi\in D(N^{1/2})$):
%----
 \be\label{N-a(k)}  \Vert N^{1/2}   \varphi  \Vert ^2   = \int_{\R^3}
 \Vert  a(k)  \varphi \Vert ^2   d k. \ee
 %----

 \section{Pull Through Formula}

Let $U_g$  be a normalized ground state given by Theorem \ref{t-gr-state}.
It is recalled in Theorem  \ref{t-gr-state} that $U_g$ belongs to the domain  $ N^m \otimes I$ for all integers  $m$.
According to Proposition \ref{a-k}, the function $k \rightarrow (a(k)\otimes I )  U_g$
  is well defined as an element of  $L^2(\R^3, {\cal H}^3 )$.
 In particular, this function is defined almost everywhere. We give in the next result an explicit expression of this function, recalling the  Pull through formula and its proof (see  \cite{C-G} and see also 
  \cite{SCH,G-J-71,F-73,G-polaron,G-J,BFS}$,\dots$). The uniqueness of the ground state is not useful for that purpose.

\begin{theo}\label{rec-0} Let  $U_g$  be a (normalized) ground state of Theorem \ref{t-gr-state}. 
Then, for almost every $k\in \R^3 \setminus \{ 0 \}$:
%---
\be\label{expr-a(k)}  (a(k)\otimes I )  U_g   = -(g/\sqrt {2}) \sum _{\lambda =1}^P    \sum _{m=1}^3
 B_{m , x_{\lambda}}(k)  ( H(g)-E(g) + |k|) ^{-1} (I  \otimes \sigma_m ^{[\lambda]}  )U_g    \ee
%---
where  $B_{m , x_{\lambda}}(k)$  is defined in (\ref{7.3}). Consequently:
%----
\be\label{majo-a(k)}  \Vert (a(k)\otimes I )  U_g \Vert  \leq (g/\sqrt {2}) \sum _{\lambda =1}^P    \sum _{m=1}^3
 \frac { |B_{m , x_{\lambda}}(k) |} {|k|}  \Vert  (I  \otimes \sigma_m ^{[\lambda]}  )U_g \Vert.    \ee
%---

\end{theo}

{\it Proof.}  In view of Lemma 2.5 in \cite{DG}:
%---
\be\label{commut} [  d \Gamma (M _{\omega }) , a(k) ] = -  |k| a(k).   \ee
%---
Thus, from (\ref{7.1}) and (\ref{7.2}), for almost every  $k$ in $\R^3$:
%----
$$ ( H(g)-E(g) + |k|) (a(k)\otimes I )  U_g  =  (a(k)\otimes I ) ( H(g)-E(g)) U_g +
 g [H_{int}, (a(k)\otimes I ) ]  U_g. $$
%---
One has:
 %-----
$$  [H_{int}, (a(k)\otimes I ) ] =  \sum _{\lambda =1}^P    \sum _{m=1}^3
   [ B_{m}(x_{\lambda})   , a(k)]  \otimes \sigma_m ^{[\lambda]}. $$
%----
Besides:
%---
\be\label{commut-B}  \sqrt {2}  [ B_m (x_{\lambda}) , a(k) ] =  [ a^{\star}(B_{m , x_{\lambda}}) , a(k)  ]
=-   B_{m , x_{\lambda}}(k). \ee
%------
Consequently:
%---
$$ [H_{int}, (a(k)\otimes I ) ] = -(1/\sqrt {2})  \sum _{\lambda =1}^P    \sum _{m=1}^3   B_{m , x_{\lambda}}(k)
 (I  \otimes \sigma_m ^{[\lambda]}  ). $$
For all $k\not = 0$, the operator $( H(g) -E (g) + |k|)$ is invertible.
One then deduces equality (\ref{expr-a(k)}).
\fpr

\end{appendices}

laurent.amour@univ-reims.fr\newline
LMR FRE CNRS 2011, Universit\'e de Reims Champagne-Ardenne\\
 Moulin de la Housse, BP 1039,
 51687 REIMS Cedex 2, France.

lisette.jager@univ-reims.fr\newline
LMR FRE CNRS 2011, Universit\'e de Reims Champagne-Ardenne\\
 Moulin de la Housse, BP 1039,
 51687 REIMS Cedex 2, France.

jean.nourrigat@univ-reims.fr\newline
LMR FRE CNRS 2011, Universit\'e de Reims Champagne-Ardenne\\
 Moulin de la Housse, BP 1039,
 51687 REIMS Cedex 2, France.

\end{document}